\documentclass[aps,prb,twocolumn,groupedaddress]{revtex4}
\usepackage[latin1]{inputenc}
\usepackage[english]{babel}
\usepackage{setspace}  
\usepackage{graphicx}
\usepackage{amssymb}
\usepackage{amscd}
\usepackage{rotating}
\begin{document}

\preprint{}

\title{Phases of lattice hard core bosons in a periodic superlattice}

\author{Long Dang and Massimo Boninsegni}
\affiliation{Department of Physics, University of Alberta, Edmonton, Alberta, Canada, T6G 2J1}
\date{\today}
\begin{abstract}
We study by Quantum Monte Carlo simulations the phase diagram of lattice hard core bosons with nearest-neighbour repulsive interactions, in the presence of a super-lattice of adsorption sites. 
For a moderate adsorption strength, the system forms crystal phases registered with the adsorption lattice; a ``supersolid" phase exists, on both the vacancy and interstitial sides, whereas at commensuration the superfluid density vanishes. The possible relevance of these results to experiments on $^4$He films adsorbed on graphite is discussed.  
\end{abstract}

\maketitle
\section{Introduction}

Films of $^4$He adsorbed on strongly attractive substrates such as graphite, have been the subject of intense experimental and theoretical investigation, motivated by the remarkable variety of phases that such films display.\cite{bruch,greywall} Although this research subject is relatively old, it has recently enjoyed a resurgence of interest in the context of a search for a {\it supersolid} phase of helium, namely one displaying simultaneously crystalline order and frictionless flow. That such a phase might exist in the second layer of an adsorbed $^4$He film on graphite, was first suggested by Crowell and Reppy over a decade ago,\cite{crowell} but this contention has been recently brought back to fore.\cite{fuku,saunders0} In particular, it is suggested that a supersolid phase may occur in the vicinity (or in correspondence) of a crystalline phase of the second adsorbed helium layer, {\it registered} with the underlying substrate. The most recent, first principles numerical studies of helium films on graphite, have yielded no evidence of such a supersolid phase, as no such registered crystal is observed.\cite{massimo08} Nonetheless, the general issue of interplay between boson localization, possibly induced by an external pinning potential, and superfluidity, remains one of general interest in condensed matter and quantum many-body physics. Moreover, theoretical predictions may also soon be tested experimentally, possibly in a more controlled fashion, with ultracold atoms in optical lattices.\cite{group4}
\begin{figure}  
\includegraphics[scale=0.25,angle=0] {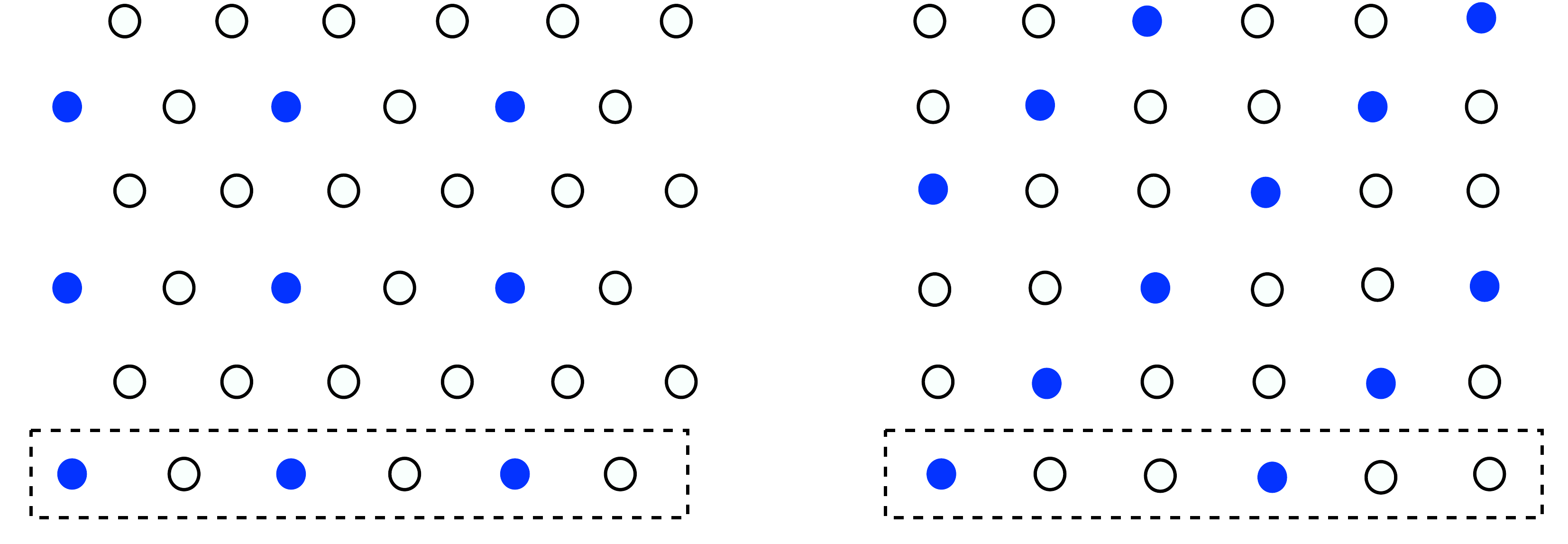}
\caption{(Color online).  Structure of the pinning potential on the triangular (left) and square (right) lattices. Filled circles represent lattice sites at which the pinning potential is worth $-\epsilon$.}
\label{epot}
\end{figure} 

That an external potential can significantly alter the phase diagram of a system of interacting bosons, giving rise to additional phases, is well known. For example, we have recently shown how a disordering potential can give rise to a glassy phase, as well as induce superfluidity in systems that do not display it, in the absence of disorder.\cite{dis}

In this work, we investigate theoretically, by means of Monte Carlo simulations, the possible existence of  supersolid phases of  many-boson systems in the vicinity of crystalline phases stabilized by external periodic potentials (such as that due to the adsorption sites of a corrugated substrate), i.e., {\it not} present in the phase diagram of the system in the absence of an external potential. Our study is based on a lattice model of interacting bosons, similar to that which has been utilized in previous theoretical works,\cite{zimmerli94} as a minimal model of the very nearly two-dimensional (2D) first  few $^4$He adlayers on graphite.\cite{campbell72,zimmerli92}  

Specifically, we consider here the (hard core) limit of infinite on-site repulsion, i.e., no more than one boson per site, and include a repulsive interaction (of strength $V$) between particles occupying nearest-neighbouring lattice sites. 
For sufficiently strong nearest-neighbour repulsion, the ground state of the system is known to be a crystal, at  particle density $\rho$=1/2 (``checkerboard" solid) on a square lattice,\cite{batrouni00} and $\rho$=1/3 (2/3) on a triangular lattice.\cite{group3}  We include here a sublattice of {\it attractive} sites as well, acting as a strong pinning potential (see Fig. \ref{epot}). The sublattice is purposefully chosen {\it not} to correspond to any  crystal structure which the system forms in the absence of an external potential. For  sufficiently strong adsorption, additional crystalline phases can be expected to appear, registered with the adsorption sublattice; henceforth, we shall refer to these crystalline phases as {\it commensurate}, the term {\it incommensurate} being used instead to refer to the solid phases that the system forms in the absence of any external potential. For example, our choice of pinning potential on the triangular lattice is such that particle density $\rho$=$\rho_C$=1/4 corresponds to a commensurate crystal, while $\rho$=$\rho_I$=1/3  to an incommensurate one. In other words, here the terms ``commensurate" and ``incommensurate" are with respect to the pinning potential.

The purpose of this study is to provide a simple theoretical framework to interpret experimental studies probing for possible (commensurate) supersolid phases of helium films on graphite.  Although we mostly discuss here numerical results obtained on a triangular lattice geometry, we have observed the same general physical behaviour on the square lattice as well.

Our main findings is that supersolid phases exist on both the interstitial and on the vacancy side of a commensurate (registered) crystal. However, the superfluid density {\it always vanishes} as the density hits a value corresponding to either a commensurate or incommensurate crystal.  In this sense, the pinning potential does not give rise to fundamentally new behaviour, with respect to what is observed in this model near and/or at incommensurate crystal phases, in the absence of any external potential.\cite{batrouni00,group3} The vanishing of the superfluid response at crystal density, appears therefore to be a general hallmark of any phase labelled as ``supersolid", occurring in a system of this type, i.e., in the presence of an external pinning potential. 

This paper is organized as follows. In Sec. \ref{mod}, we describe the model Hamiltonian. In Sec. \ref{mt}, we briefly review the methodology adopted in this work, while our 
numerical results are presented in Sec. \ref{res}. Finally, we outline our conclusions in Sec. \ref{con}.
\section{Model}\label{mod}
The 2D hard core Bose model with nearest-neighbour interactions is expressed as follows:
\begin{equation}
H= -t\sum_{\langle ij\rangle}(\hat{a}_i^{\dagger}\hat{a}_j + h.c.) +  V\sum_{\langle ij\rangle}\hat{n}_i\hat{n}_j -
  \sum_i  \mu_i  \hat{n}_i\ . \label{ham}
\end{equation}

We consider here a lattice (either triangular or square) of $N=L\times L$ sites, with periodic boundary conditions. The sums $\langle ij\rangle$ run over all pairs of nearest-neighbouring lattice sites, $\hat a^\dagger_i$ $(\hat a_i)$  is the Bose creation (annihilation) operator for a particle at site $i$, $\hat{n}_i=\hat{a}_i^{\dagger}\hat{a}_i$ is the local density operator, $t$ is the hopping amplitude, $V$ is the strength of nearest-neighbour repulsion, while $\mu_i = \mu-\epsilon_i$ is a site-dependent chemical potential. Here, $\epsilon_i=\epsilon$ if the site belongs to the pinning sublattice (see Fig. $\ref{epot}$), zero otherwise, $\epsilon > 0$ being the strength of the pinning potential.   As mentioned above, a {\it hard-core} on-site repulsion is assumed, limiting the occupation of every site to no more than one particle.

The pinning potentials have been chosen for definiteness to correspond to commensurate density $\rho_C$=1/4 for the triangular lattice,  $\rho_C$=1/3 for the square lattice (Fig. $\ref{epot}$). No particular physical significance should be ascribed to these choices, motivated only by the goal of making  commensurate phases lower in density than the incommensurate ones, as would be the case for the second layer of $^4$He on graphite, if a commensurate crystal exists. It seems reasonable to expect that the basic physical conclusions ought to remain unaffected by a different choice of pinning sublattice.

\section{Methodology}\label{mt}
We perform grand-canonical quantum Monte Carlo simulations to study the ground state properties of (\ref{ham}), using the Worm Algorithm in the lattice path-integral representation. As the details of this computational method are extensively described elsewhere,\cite{prokofiev98,lode07} and because the calculations performed here are standard, we shall not review it here, and simply refer interested readers to the original references.

The results shown here correspond to a temperature $T$ sufficiently low (typically $\beta=1/T\approx L$), so as to be regarded as essentially ground state estimates. 
In order to characterize the various phases, we compute the superfluid fraction $\rho_s$, as well as the static structure factor:
\begin{equation}
S({\bf Q})= \frac{1}{N^2}\biggl \langle \biggl | \sum_{i=1}^{N}\hat{n}_i e^{i {\bf Q.r_i}}\biggr |^2 \biggr \rangle
\end{equation}
where $<...>$ stands for thermal average. The presence of crystalline long-range order is signaled by a finite value of $S({\bf Q})$ for some specific wave vector, in the thermodynamic limit. For the triangular lattice, ${\bf Q} = (\pi,{2\pi}/{\sqrt{3}})$ is a wave vector corresponding to the registered (commensurate) crystal at $\rho$=$\rho_C$=1/4, while ${\bf Q} = ({4\pi}/{3}, 0)$ to an incommensurate crystal with $\rho$=$\rho_I$=1/3, 2/3. On the square lattice, ${\bf Q}=({4\pi}/{3},-{2\pi}/{3})$ corresponds to a registered crystal at $\rho$=$\rho_C$=1/3, and ${\bf Q}=(\pi, \pi)$ to an incommensurate (checkerboard) crystal at $\rho$=$\rho_I$=1/2.

Unless otherwise specified, the results discussed below pertain to a triangular lattice geometry.  We have carried out careful extrapolation of the results to the thermodynamic limit.  In general, we observed that  estimates obtained on a lattice of $L\times L$=576 sites  are identical, within statistical uncertainties, with the extrapolated ones. 

\section{Results}\label{res}


\begin{figure}[!t]
\includegraphics[scale=0.25,angle=0]{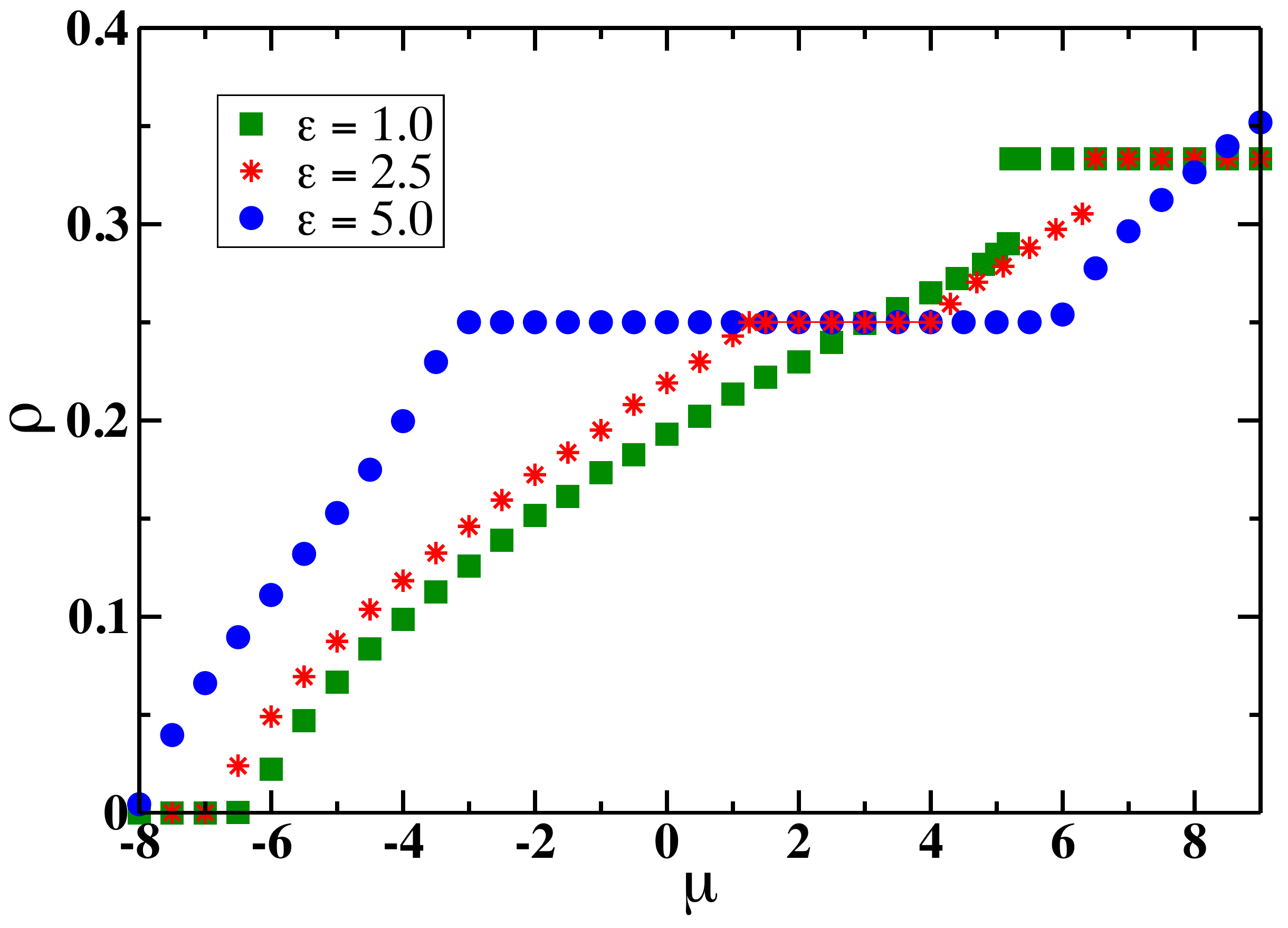}
\caption{(Color online). Density $\rho$ versus chemical potential $\mu$ for $V$=6 and three different pinning potential strengths,  namely $\epsilon$=1 (filled squares), $\epsilon$=2.5 (stars) and $\epsilon$=5 (filled circles). Statistical errors are smaller than symbol sizes. Results shown are for a triangular lattice with $L$=24.}
\label{V6}
\end{figure}

We begin by discussing the existence of registered (commensurate) solid phases for sufficiently large values of $\epsilon$, i.e., the strength of the adsorption potential.
Fig. \ref{V6} displays the density computed as a function of the chemical potential for $V = 6$, for three different pinning potential strengths,  namely $\epsilon$=1 (filled squares), $\epsilon$=2.5 (stars) and $\epsilon$=5 (filled circles). The value of $V$ is large enough for the incommensurate solid phases to exist, in the model without pinning potential. 

For a weak pinning potential, plateaus in the density appear only at $\rho=\rho_I=1/3$ and $\rho$=2/3, i.e., in correspondence of the incommensurate phases. A discontinuity of the curve signals a first-order phase transition between a superfluid and the incommensurate crystal.  Analogously to what observed in the model without pinning potential,\cite{group3} for sufficiently large $V$ the $\rho(\mu)$ curve is continuous on the interstitial side ($\rho >\rho_I$), as a supersolid phase exists. 

As the strength of the pinning potential is increased, two additional crystalline phases appear, one at $\rho=\rho_C=1/4$, the other at $\rho$=5/8. Henceforth, we shall focus our attention on the phase of density $\rho_C$, which is registered with the pinning potential. The other phase arises from the competition between the pinning potential and the nearest-neighbour repulsion, and the basic physics at or near this density is the same as near $\rho_C$. As shown in Fig. \ref{V6}, the $\rho(\mu)$ curve displays no discontinuities on either the vacancy or the interstitial side of the commensurate crystal. This is evidence of vacancy- and interstitial-doped supersolids, as we discuss below. 

\begin{figure}
\includegraphics[scale=0.35,angle=0]{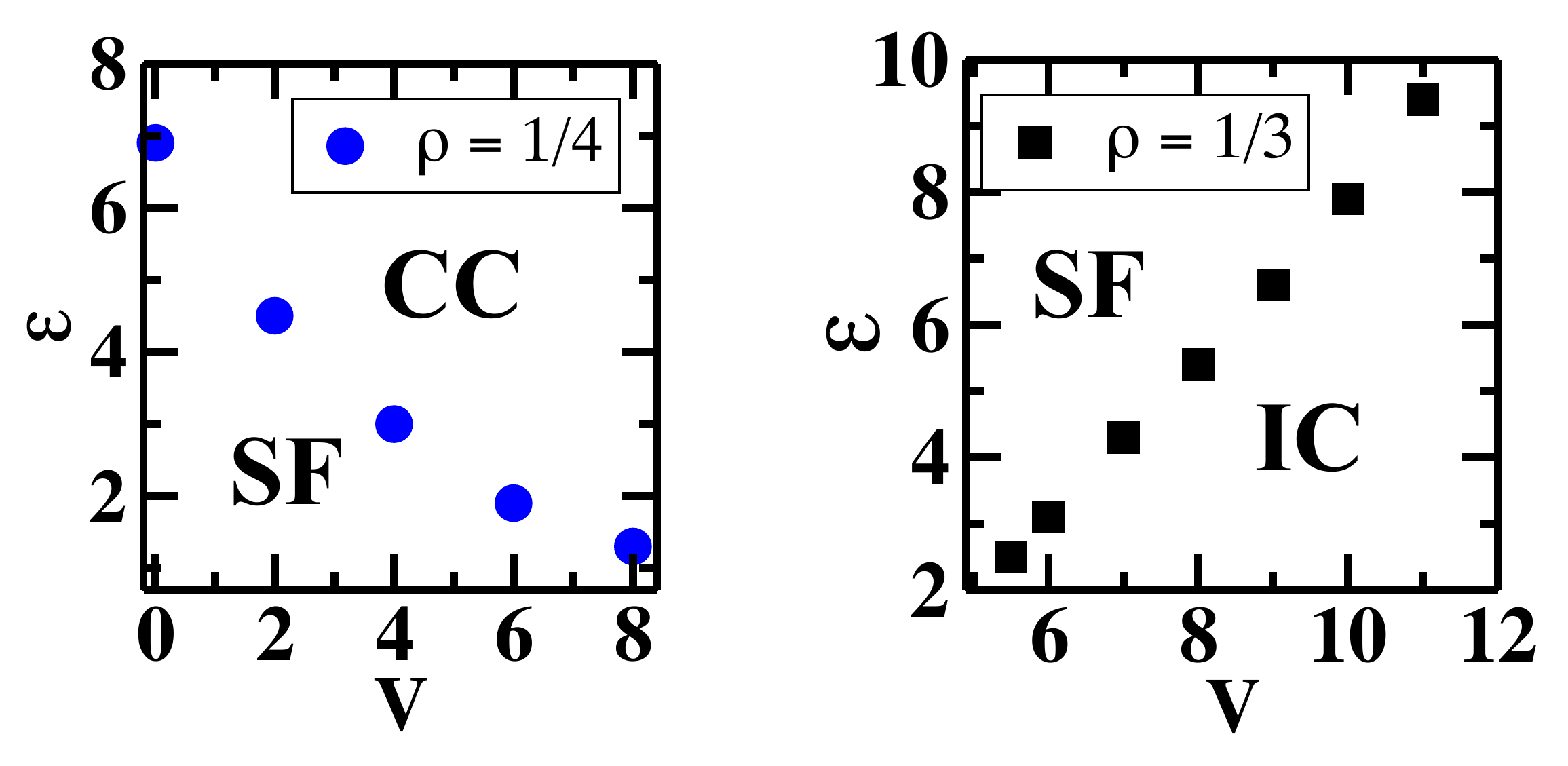}
\caption{(Color online). Ground state phase diagram of (\ref{ham}). Symbols lie at computed phase 
boundaries between a superfluid (SF) and a commensurate crystal (CC) for  $\rho=1/4$ (left), and between a superfluid and an incommensurate crystal (IC) 
at  $\rho=1/3$  (right).
Statistical errors are smaller than symbol sizes.}
\label{phasediagram}
\end{figure}

On performing a sufficient number of runs in the $(V,\epsilon)$ plane, we have computed the phase boundary lines shown in Fig. \ref{phasediagram}, between a superfluid and crystal at the two densities $\rho_C$ and $\rho_I$. The left part of the figure refers to $\rho_C$. The system is superfluid for $\epsilon < \epsilon_c(V)$, where $\epsilon_c(V)$ is the minimum strength of the pinning potential for which a commensurate crystal is present, as a function of the strength of the nearest-neighbour repulsion $V$. For the commensurate phase $\epsilon_c(V)$ is monotonically decreasing with $V$, as the presence of a strong nearest-neighbour repulsion, which causes the appearance of the incommensurate crystalline phase at $\rho_I=1/3$ also favours  the formation of a commensurate crystal at $\rho_C$ (in fact, $\epsilon_c(V)$ approaches zero as $V\to\infty$). On the other hand, the right part of Fig.  \ref{phasediagram} shows that the pinning potential {\it suppresses} crystallization at density $\rho_I$, i.e., a {\it greater value} of $V$ is needed to stabilize the incommensurate crystal at $\rho_I=1/3$ if the external pinning potential is present. This is due to  the lattice mismatch of two competing crystalline phases. A sufficiently large value of $\epsilon$ causes  the incommensurate phase to disappear altogether.

\begin{figure}[!t]
\includegraphics[scale=0.28,angle=0]{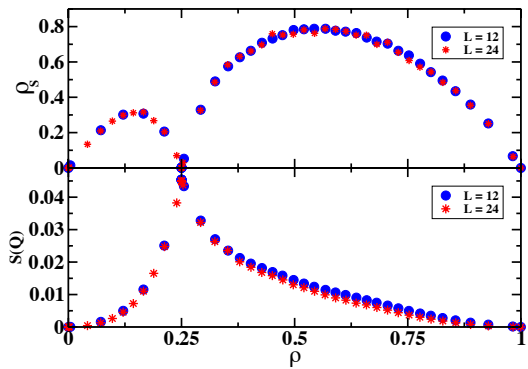}
\caption{(Color online). Superfluid density $\rho_S$ (upper panel) and static structure factor S(${\bf Q}$) (lower panel) in case of V = 4.0, $\epsilon = 5.0$ and a wave vector ${\bf Q}=(\pi,{2\pi}/{\sqrt{3}})$. Statistical errors are smaller than symbol sizes.}
\label{e5V4sf}
\end{figure}

We now discuss the superfluid properties of the system near crystallization. We begin by examining the physics of the system near a commensurate solid phase. Fig. \ref{e5V4sf} shows the superfluid fraction $\rho_S$  (upper panel) and the static structure factor $S({\bf Q}=(\pi,{2\pi}/{\sqrt{3}}))$  as a function of the particle density. The choice of parameters, namely $V$=4 and $\epsilon$=5, corresponds to a situation in which the only crystalline phase that the system forms is the commensurate one, at a density $\rho_C$. 
\hfil\break
Both $\rho_S$ and $S({\bf Q})$ are everywhere {\it finite}, except at exactly $\rho_C$ where the superfluid response vanishes. The fact that $\rho(\mu)$ is continuous everywhere, allows one to rule out coexistence of two phases (superfluid and crystal) possessing only one of the two types of order. Thus, based on its strict definition, one would have to conclude that this system is everywhere ``supersolid", except at commensurate density. However, such a denomination appears to be meaningful (if at all) only in the vicinity of $\rho_C$, where the physical character of the phase can be surmised to be that of a commensurate crystal doped with either vacancies or interstitials. Away from $\rho_C$, the nature of the system is basically that of a fluid with a density modulation arising from the pinning potential.\cite{note}

That at {\it exactly} $\rho=\rho_C$ the superfluid fraction vanishes, is a significant result that warrants a few comments.
 Supersolidity in model (\ref{ham}) on the triangular lattice (it is not present on the square lattice), requires that a crystal be doped with interstitial particles, i.e., the superfluid density of an undoped crystal is always zero.\cite{group3} However, one might speculate that the lower density commensurate phase stabilized by the external potential might enjoy different properties than the incommensurate one, which is the only one observed in the absence of a pinning potential. 
We find, however,  that the superfluid density always vanishes at $\rho_C$, on both lattice geometries considered here. We have also repeated the same analysis for different choices of the parameters, including those for which both commensurate and incommensurate phases exists, but the presence of an incommensurate phase at higher density does not alter the physics of the system in the vicinity of $\rho_C$, i.e., a ``supersolid" phase exists on both the interstitial and vacancy sides, but not at commensuration.

This is a result of potential experimental relevance, as studies of adsorbed $^4$He films on corrugated substrate on graphite, for which claims of possible supersolid behaviour near commensurate density are made, can determine the superfluid response as a function of coverage. It appears from our results that, to the extent that (\ref{ham}) can be regarded as a reasonable qualitative model of a thin helium film on a corrugated substrate, the superfluid signal must vanish at the  coverage  corresponding the occurrence of a commensurate crystal, if one is to make a claim of a ``supersolid" phase near or at commensuration.  

The physics of the system near the incommensurate crystal phase is the same as in the absence of an external potential.\cite{group3} In particular, the superfluid density again always vanishes at $\rho_I$.  Here too, one might have expected that the weakening of the incommensurate crystal caused by the pinning potential could give rise to a ``softer" crystalline phase, capable of superflow. What is observed, however, is that as long as the incommensurate crystal exists, the superfluid density at $\rho_I$ vanishes.  There is always a first-order phase transition on the vacancy side; on the other hand, on the interstitial side, depending on the value of $V$ one may have a first-order phase transition to a superfluid or a second-order phase transition to a supersolid. 

\begin{figure}[!t]
\includegraphics[scale=0.28,angle=0]{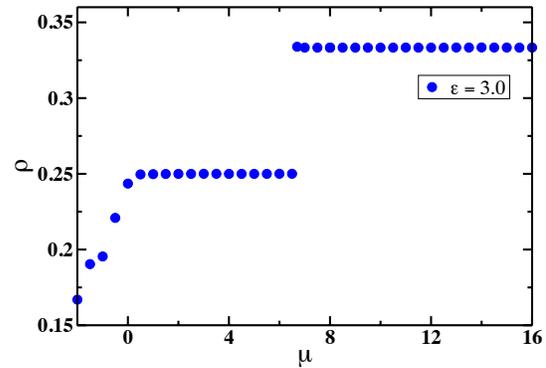}
\caption{(Color online). Density $\rho$ versus chemical potential $\mu$ for the ground state of (\ref{ham}), with V = 8, $\epsilon = 3$. The density jump signals a first-order phase transition between a commensurate and an incommensurate crystal.}
\label{crycry}
\end{figure}

We now consider the regime in which both the commensurate and incommensurate phases exist, and explore the quantum phase transitions between the two, with the possible occurrence of intervening phases. There are two possible scenarios which have been observed: one is a direct transition from the commensurate to the incommensurate crystal, through a first order transition. In other words, there is only a jump in the curve $\rho(\mu)$ from $\rho_C$ to $\rho_I$, as shown in Fig. \ref{crycry}. This occurs roughly in a regime where $\epsilon << V$, and is a scenario that appears to apply to the first layer of helium on graphite,\cite{massimo08} or to films of molecular hydrogen adsorbed on graphite,  or other corrugated substrates.\cite{turnbull}
\\
The other scenario, occurring for $\epsilon \sim V$, is a second-order transition from the commensurate crystal at $\rho_C$ to a doped supersolid phase, and then, as the density is increased, to a superfluid, followed by a first order phase transition from the superfluid to the incommensurate crystal at $\rho_I$. The transition from supersolid to superfluid is indicated by the change in slope of the $\rho(\mu)$ curve. Again, we should note that this superfluid phase with a density modulation arising from the pinning potential. 

\section{Conclusions}\label{con}
We have studied the ground state phase diagram of lattice hard core lattice bosons with a nearest-neighbour repulsion, in the presence of an external periodic potential mimicking a superlattice of adsorption sites. We have utilized an exact numerical method. Our goal, in using such a simplified model, was obviously not that of achieving a realistic description of an actual adsorbed film on a given substrate, but rather to gain qualitative understanding of the different phases that one may be able to observe experimentally. We have carried out our studies on triangular and square lattice geometries, as well as different choices of external potential periodicity, always chosen to be {\it not} commensurate with the crystalline phase that the system forms in the absence of a potential. The main findings are independent of the lattice geometry and/or potential periodicity.

Our results show that in specific circumstances ``supersolid" phases (whether or not such terminology is appropriate, given the crucial role played by the external potential, is a matter of debate) can exist in the vicinity of commensurate crystals stabilized by the adsorption potential. A distinctive signature of the occurrence of such phases is the vanishing of the superfluid density at commensuration. This seems to be a universal feature of this type of system, one that we would expect to see in any experiment claiming observation of a supersolid phase of adsorbed films (e.g., of $^4$He) on substrates such as graphite.
It is also worth mentioning that these predictions may also lend themselves to possible experimental verification by means of ultracold atoms in optical lattices.\cite{group4}

\section*{Acknowledgments}
This work was supported in part by the Natural Science and Engineering Research Council of Canada under research grant G121210893, by the Alberta Informatics Circle of Research Excellence and by the Swiss National Science Foundation. Part of the simulations ran on the Hreidar cluster at ETH Zurich. We would like to thank Lode Pollet, Piyush Jain and Joseph Turnbull for fruitful discussions. 


\begin{thebibliography}{99}
\bibitem{note} There is clearly an important conceptual difference between a physical system in which the breaking of translational invariance occurs spontaneously, and one in which it is induced by an external potential. Indeed, the very use of the terminology ``supersolid" in the latter case may be questionable.
\end{thebibliography}
\end{document}